\documentclass[runningheads]{llncs}
\usepackage{bm}
\usepackage{amsmath}
\usepackage{amssymb}
\usepackage{graphicx}
\usepackage{physics}
\usepackage[utf8]{inputenc}
\usepackage[T1]{fontenc}
\usepackage{float}
\usepackage{braket}
\usepackage[table,xcdraw]{xcolor}
\usepackage{tikz}
\usepackage{subfigure}
\begin{document}
\title{Solving Combinatorial Optimization Problems on a Photonic Quantum Computer}
%
%
\author{Mateusz Slysz\inst{1, 2}\orcidID{0000-0003-3124-9899} \and
Krzysztof Kurowski\inst{1}\orcidID{0000-0002-4478-6119} \and
Grzegorz Waligóra\inst{2}\orcidID{0000-0003-2108-1113}}

\authorrunning{M. Slysz et al.}
\titlerunning{Combinatorial Optimization on Photonic Quantum Computer}
%
\institute{Poznań Supercomputing and Networking Center, IBCH PAS\\
\email{\{mslysz, krzysztof.kurowski\}@man.poznan.pl}
\and
Poznań University of Technology\\Institute of Computing Science\\Poznań, Poland\\
\email{\{grzegorz.waligora\}@cs.put.poznan.pl}
}
\maketitle              
\begin{abstract}
Combinatorial optimization problems pose significant computational challenges across various fields, from logistics to cryptography. Traditional computational methods often struggle with their exponential complexity, motivating exploration into alternative paradigms such as quantum computing. In this paper, we investigate the application of photonic quantum computing to solve combinatorial optimization problems. Leveraging the principles of quantum mechanics, we demonstrate how photonic quantum computers can efficiently explore solution spaces and identify optimal solutions for a range of combinatorial problems. We provide an overview of quantum algorithms tailored for combinatorial optimization for different quantum architectures (boson sampling, quantum annealing and gate-based quantum computing). Additionally, we discuss the advantages and challenges of implementing those algorithms on photonic quantum hardware. Through experiments run on an 8-qumode photonic quantum device, as well as numerical simulations, we evaluate the performance of photonic quantum computers in solving representative combinatorial optimization problems, such as the Max-Cut problem and the Job Shop Scheduling Problem.

\end{abstract}
\section{Introduction}
Quantum computing is a rapidly growing field of science, thanks to recent technological advances, as well as strong interest from both the scientific community and the tech industry, which are looking for potential applications of this groundbreaking technology.
Quantum computers are being produced, by both tech-giants and smaller start-ups, but the multitude of technical solutions on the market makes it impossible to clearly state which of the current technologies will be the leading one, in the context of future large-scale production of quantum devices.
Popular architectures include the gate-based model - a universal model of quantum computing - and quantum annealing (QA) which is useful for solving a certain class of optimization problems.
In addition to various architectures there are also many different physical implementations of those quantum computing paradigms.
For the moment, the largest and most popular quantum computing vendors such as IBM and D-Wave are using superconducting qubits, which have many advances, including scalability and fast computation times.
However, the main disadvantage of this technology is that qubits need to be cooled down to extremely low temperatures, close to absolute zero, which is both technically difficult and very expensive to maintain, making the widespread installation of those devices in non-specialized facilities impossible in practice.

Consequently, other paradigms of quantum computing are still being sought. One of the more interesting ideas is to use optical components to build a quantum device.
Working with photons would solve the cooling problem, since such systems are capable of operating at room temperature.
In addition, they have many other advantages - thanks to the widespread production of optical components and maturity of technical solutions in this field, it would be possible to easily develop, install and service those devices.
Also, it is possible to integrate with quantum communication systems, with no need for the complex process of conversion to the optical domain and photon wavelength conversion.

In this paper we aim to test and benchmark the capabilities of such a photonic quantum device. Our tests are based on a certain class of Quadratic Unconstrained Binary Optimization (QUBO) problems. We describe hybrid classical-quantum algorithms, that allow solving such problems on a photonic quantum device, as well as compare their performance with classical simulations and other quantum architectures.

This paper is organised as follows: in Section 2 we describe the principles of photonic quantum computers based on the boson sampling paradigm, along with of a specific physical implementation of such a machine used in our experiments. In Section 3 we describe different optimization algorithms for different quantum computer architectures, emphasizing the description of the Binary Bosonic Solver algorithm, used for optimization on bosonic machines. In Section 4, we formulate two discrete optimization problems: Max-Cut and Job-Shop Scheduling Problem, and then in Section 5 we conduct computational experiments using the instances of these problems on both classical and quantum computers. Section 6 contains conclusions from the obtained experimental results and potential ideas for future work and improvements.


\section{Photonic Quantum Computer}
\subsection{Boson Sampling}
The described photonic quantum computer is an implementation of a computation technique called Boson Sampling \cite{PhysRevLett.119.170501}. Boson Sampling (BS) is a quantum computing paradigm that takes advantage of quantum phenomena occurring between bosons. Photons belong to the class of boson particles, which makes it possible to create a quantum processor based on an optical circuit with a single photon source and a single photon detector.
By creating a maze of optical paths using beam-splitters, we can create quantum states, with amplitudes correlated to the beam-splitter reflectance and transmittance parameters.
Every time a photon passes through such an optical intersection, a superposition is produced.
Also each meeting point of two or more photon paths produces an entangled state.
The large number of potential paths and branches through which particles can travel gives the possibility of obtaining very large entangled states, even for a small number of photons, which determines the quantum advantage of such a device over classical simulation.

The readout of the result involves sampling the probability distribution by measuring the number of photons in each detector at the system's output.
By adjusting the parameters of optical gates, we can make the desired output results more likely.
The difference with other quantum computing devices is that measurements in the BS process are made in the photon number domain and can take integer values that do not directly correspond to qubit measurements. Therefore, a better way to describe this process is to use the concept of quantum modes or "qumodes" instead of qubits.

\begin{figure}[H]
    \centering
    \includegraphics[width = 0.8\textwidth]{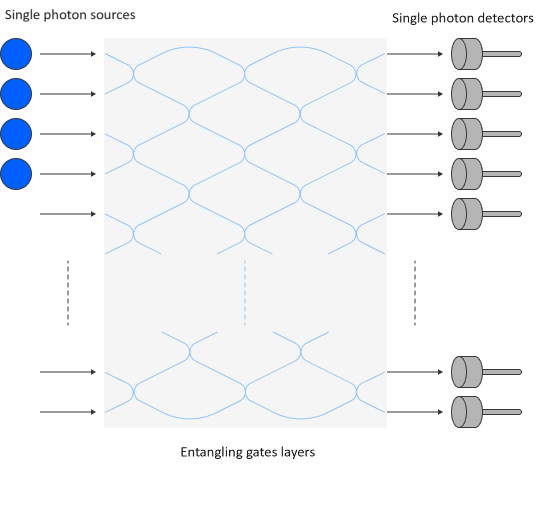}
    \caption{A schematics of a photonic quantum computer.}
    \label{fig:photonic_qc}
\end{figure}

\subsection{ORCA PT-1}
In this research, we are going to use an ORCA PT-1 photonic quantum computer \cite{ORCA_shallow}, installed at Poznan Supercomputing and Networking Center.
The PT-1 systems have been constructed in a way to reduce the number of optical components in the circuit. With only one single photon source and one single photon detector, it is possible to create a BS quantum device by shifting calculations from the spatial domain to the time domain.
This interferometer consists of a series of optical loops with programmable beam-splitters at the entrance to each one. 
Each time a photon approaches such a loop, it can stay in it for another time interval, or move on, with a probability that depends on the parameters of the beam-splitter.
This creates an entangled state consisting of a superposition of different photons in the output time slots. The system performs a BS process by sending a sequence of individual photons at subsequent time intervals. In the case of the ORCA PT-1 system, the qumodes are defined by time intervals. This architecture is equivalent to a spatial-domain boson sampler with gates between each pair of neighbouring qumodes.

\begin{figure}[H]
    \centering
    \includegraphics[width = \textwidth]{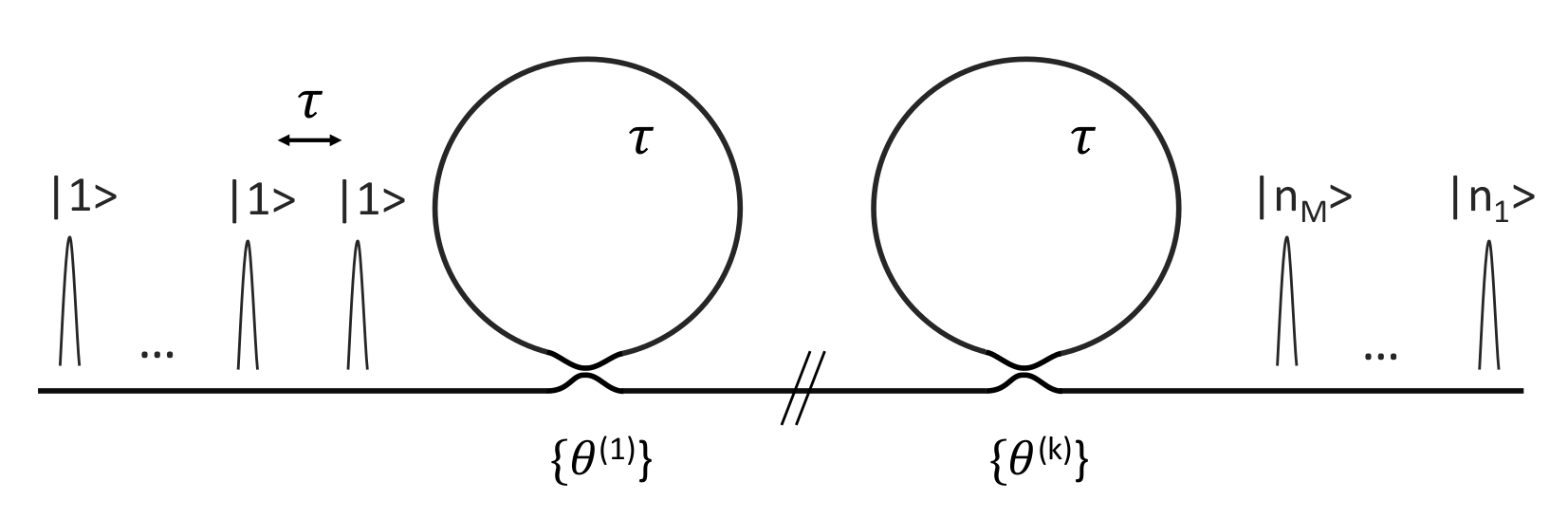}
    \caption{A schematics of an ORCA PT-1 photonic quantum computer.}
    \label{fig:orca_tbi}
\end{figure}

\subsection{Double-loop interferometer}
\begin{figure}
  \centering
  \subfigure[Single-loop interferometer]{\label{fig:I1}\includegraphics[width=0.45\textwidth]{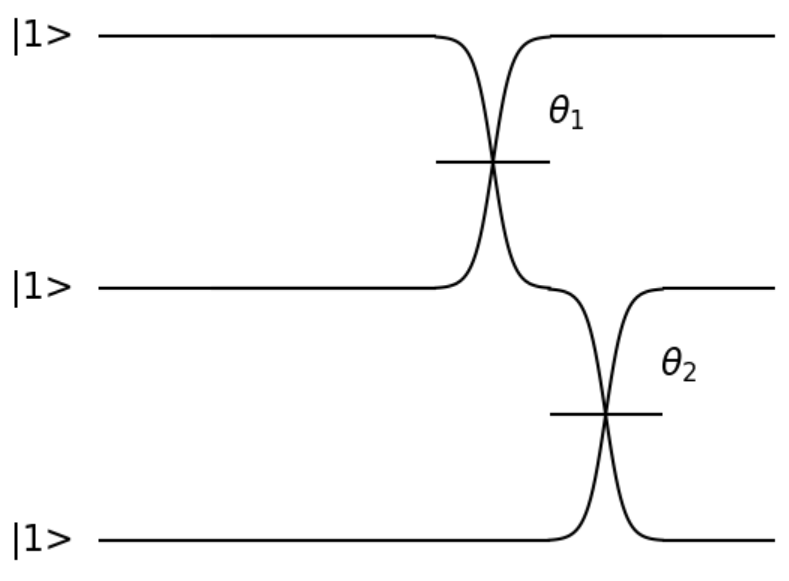}}\qquad
  \subfigure[Double-loop interferometer]{\label{fig:I2}\includegraphics[width=0.45\textwidth]{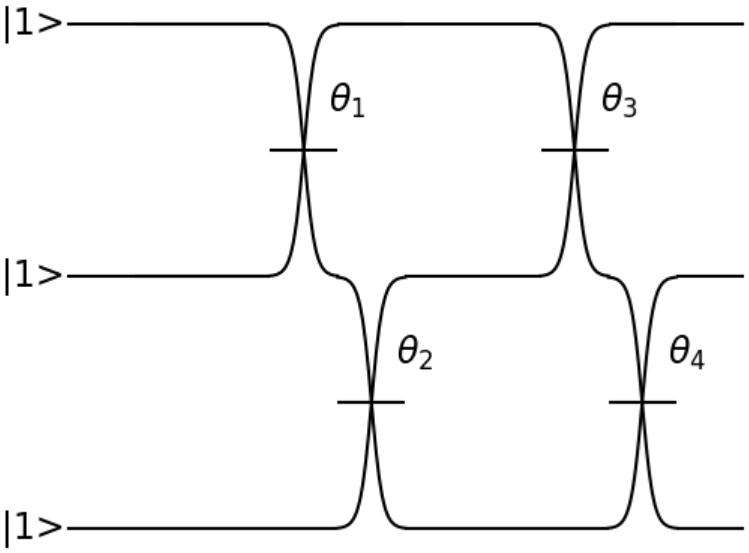}}
\caption{Difference between single-loop and double-loop interferometer. The double-loop
inteferometer has twice as many parameters and can produce more complicated entan-
gled states.}
\label{fig:1and2loop}
\end{figure}

Another advantage of this design is the ability to easily add additional loops and control parameters to the circuit. The PT-1 device can work either in a single-loop mode, or multi-loop mode. Adding additional loops in the physical circuit is equivalent to adding an additional layer of beam-splitters between each neighbouring pair of qumodes. In the Fig. \ref{fig:1and2loop} we can see the logical equivalents for optical circuits for single and double-loop BS device. A double-loop circuit is theoretically much more difficult to simulate, making the advantage of using an actual quantum device over a simulator magnified. In general, the more loops, the exponentially more difficult it will be to simulate their operation on a classical computer.

\section{Optimization on quantum computers}

\subsection{Overview of quantum optimization methods}
Quantum computers are devices that have been proven to be exponentially faster than classical computers, which is a huge leap not only in optimization, but the whole of computer science.
For more than a few decades, there have been theoretical evidence of exponential speed-up for specific problems solved by quantum algorithms such as Deutsch's, Grover's, and Shor's \cite{nielsen2010quantum}.
However, these algorithms require a large number of fault-tolerant qubits to be used in practical applications and achieve so-called quantum advantage.
Right now, we are in a stage of development referred as the Noisy Intermediate-Scale Quantum (NISQ) era \cite{Preskill2018}, which is characterized by quantum processors which do not have sufficient number of qubits, nor are fault-tolerant enough to gain an advantage over classical computing in real-world problems.
However, there are some fields in which these kinds of computers can be used effectively, such as combinatorial optimization - a domain of finding the optimal object from a finite set.
Thanks to the certain properties of quantum computers, such as superposition and entanglement between variables, it is possible to efficiently search through the solution space, thus, quickly receive near-optimal results of a given optimization problem.
There exist several quantum computing paradigms, that approach the optimization problems in different ways.

For Quantum Annealing a certain class of Quadratic Unconstrained Binary Optimization (QUBO) \cite{Glover2022} problems can be solved using the Ising model. This happens by encoding a problem Hamiltonian on the quantum device and starting the annealing process, which is a physical process that naturally converges towards a ground state of this system, representing an optimal solution of an optimization problem. A review of heuristic approaches for the maximum cut (Max-Cut) problem and QUBO is given in \cite{dunning2018works}. Applications of QA for the Job-Shop Scheduling Problem (JSSP) can be found in \cite{kurowski2020hybrid}, \cite{Venturelli2015}, \cite{Carugno2022}, \cite{Armas2024}.

Also for the gate-based model, there exists a family of variational methods for optimizing such problems. One of the most popular ones is called Quantum Approximate Optimization Algorithm (QAOA) \cite{Fahri_QAOA}. The algorithm works by encoding the problem Hamiltonian into a parametrised variational quantum circuit, and classically optimising its parameters.
The behaviour, performance analysis and possibilities of executing QAOA for solving Max-Cut have been studied in \cite{Crooks2018}, whereas for JSSP in \cite{Amaro2022} and \cite{kurowski2023application}.

\subsection{Binary Bosonic Solver}

In this paper we want to focus on solving chosen optimization problems on a photonic quantum device. For the BS architecture there also exists a variational hybrid (quantum-classical) algorithm called Binary Bosonic Solver (BBS) \cite{ORCA_shallow}, that utilises the quantum device as a sampler. The core idea behind this algorithm is to utilise the samples collected from a BS process as candidate solution vectors and classically optimize the quantum circuit parameters to minimize the cost function of a given optimization problem.

To encode the problem, we define the $Q$ matrix, which must take into account all relationships between variables. The values at the intersection of each row and column indicate the relationship between a given pair of variables, while the main diagonal encodes the probability of selecting each variable in the final solution.

The algorithm initiates quantum gate's control parameters $\bm{\theta}$ at random and the BS process is performed, which returns an output of a set of samples (sequences of natural numbers). However, since we are considering a binary problem, the readout can be simplified to a binary value by applying a threshold, resulting in a $0$ for no photons measured and $1$ for the any positive number of photons detected. Then the cost function expression $C = \bm{x^T} Q \bm{x}$, is calculated, where the $Q$ matrix is defined specifically for the problem instance.

The cost function is minimized with respect to the quantum circuit parameters $\bm{\theta}$, by using a classical optimization algorithm, such as SPSA - a gradient-based algorithm commonly used in machine learning and continuous optimization.
New parameters $\bm{\theta'}$ are calculated and with this new set of parameters, we can now return to executing a quantum circuit on a quantum device and close the algorithm loop.
After a certain number of iterations, or after achieving a low-enough cost function value, the final result can be returned, which is a sample from the quantum circuit with optimal set of parameters $\bm{\theta^{opt}}$. The boson sampler with trained parameters should now return a vector that represents the optimal solution of a given problem with high probability.

\subsection{BBS modifications}

Due to the small number of qumodes available in currently available devices, the basic BBS algorithm is quite limiting in terms of the number of variables in the optimization problems analyzed. In order to tackle this issue, a tiling technique has been added to the algorithm to allow processing more variables than qumodes in a single run of the algorithm's loop. This allows us to divide a larger input into smaller chunks, which not only allows to fit them in, but also speeds up the computations, as jobs with a fewer number of photons tend to run a lot faster on the quantum device. In the case of chunks of uneven size, a padding is used, which allows the smaller part of the vector to be used in the last part of the calculation.

The tilling technique works by repeatedly running the boson sampling process for each new tile, and calculates gradients for each new input. Once all the samples are collected, it is possible to execute a full optimization run of the training algorithm for the concatenated output vectors. This, unfortunately, results in a quadratic increase in the number of necessary runs of the quantum circuit, but still makes it possible to perform the necessary calculations for larger instances in polynomial time.

Another problem of the current hardware is the limitation on the number of input photons to the device, due to detector calibration issues. Because of this, we must either use only a small number of available qumodes, or alternatively use sparse encoding, where, for example, only every second channel will have a photon input. To fit larger problem instances it is necessary to use the latter, however, this results in worse coverage of the search space. To compensate for this loss, additional classical trainable parameters are added to the algorithm. Each of those numbers represents a probability of a bit flip in the final solution. This means, that the number of classical parameters is equal to the number of qumodes (and the problem instance size).

\begin{figure}[H]
    \centering
    \includegraphics[width = \textwidth]{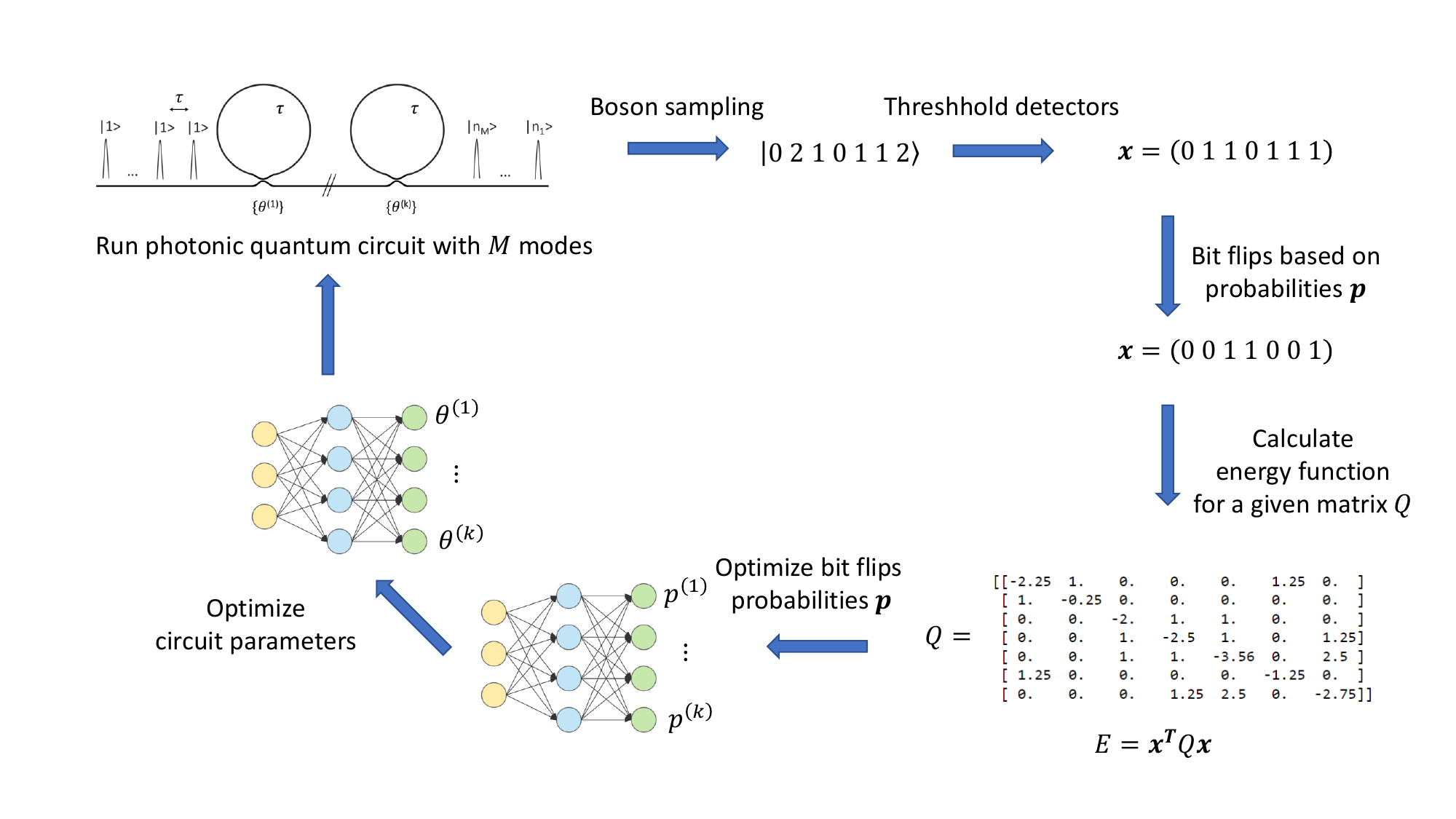}
    \caption{A schematics for a hybrid optimization algorithm which utilises a photonic quantum computer.}
    \label{fig:BBS}
\end{figure}

\section{Optimization problems}
We aim to test and benchmark the results of chosen optimization problems on a real quantum device, and compare them with the results obtained from a simulator. For this purpose we chose two combinatorial optimization problems. The first one is Max-Cut, which due to a relatively simple structure can be used as a benchmark of the quantum computer's capabilities and compared with the classical simulator of such a device. The other one is JSSP which is a more complicated optimization problem, and can showcase actual capabilities of the photonic quantum device and the BBS algorithm, as well as can be compared with other approaches that were covered on other quantum architectures.

\subsection{Max-Cut}
Max-Cut is an NP-hard problem of finding a cut in a graph, whose size is at least equal to the size of any other cut. Given a graph $G = \left(V, E\right)$ we split the vertices into 2 subsets: $V_1$ and $V_2 = V \setminus V_1$ as cutting the edges of a graph is equivalent to dividing the vertices into 2 disjoint sets.

This can be written as a binary optimization problem, where each vertex is assigned a binary variable, depending on which subset it is in. 

\begin{equation*}
    x_i = \begin{cases}
    1, & \text{if $v_i \in V_1$}.\\
    0, & \text{if $v_i \in V_2$}.
  \end{cases}
\end{equation*}

For maximizing the number of cuts one needs to maximize the expression that favours the neighbouring vertices into opposite subsets:

\begin{equation}
    \max_{\left( v_i, v_j \right) \in E} x_i + x_j - 2 x_i x_j ,
\end{equation}

\noindent or minimize the negated expression. This formulation is equivalent to solving a QUBO problem with a cost function:

\begin{equation}
    C = \bm{x^T}Q\bm{x}
\end{equation}

\noindent with the $Q$ matrix coefficients defined by the Max-Cut coefficients. The $Q$ matrix can be defined as follows:

\begin{equation*}
    Q_{i,j} = \begin{cases}
    2, & \text{if $\left (v_i, v_j \right) \in E$}.\\
    0, & \text{otherwise}.
  \end{cases}
\end{equation*}

\begin{equation*}
    Q_{i,i} = - deg (v_i) = -\sum_{ \left( v_i, v_j \right) \in E} j
\end{equation*}

\subsection{JSSP}
In JSSP a set of dedicated machines is to perform tasks of some jobs. Each job is composed of an ordered list of tasks, from among which every task requires a specific machine for a known processing time. Several constraints are imposed on jobs and machines: (i) tasks are nonpreemptable, (ii) tasks of a given job are precedence-related, whereas tasks of different jobs are independent, (iii) each task can be performed on one machine at a time, and (iv) each machine can process only one task at a time. The problem is to minimize the makespan, i.e. the maximum completion time of all tasks. JSSP belongs to the most intractable scheduling problems known in the literature, and it is NP-hard in the strong sense.

The JSSP formulation we consider is defined as follows.
There are $J$ jobs $\mathcal{J} = \set{j_1, \dots, j_J}$, each consisting of $O_j$ operations $\mathcal{O}_j = \set{O_{j1}, \dots, O_{j O_j}}$, which are supposed to be processed in a predefined order.
Each operation $O_{j,k}$ has a duration time $l_{j,k}$ and must be processed on a specified machine from a set of $M$ machines $\mathcal{M} = \set{m_1, \dots, m_M}$. A set of operations $O_{j,k}$ that have to be executed on the machine $m_m$ can be denoted as $I_m$.

The goal of JSSP is to minimize the makespan, so it seems a fairly simple optimization function for the algorithm. This time, however, we must also take into account all the constraints, related to the feasibility of the solution. For the purpose of using the quantum computer as a solver for JSSP, we need to encode the problem variables to match the QUBO notation. Inspired by \cite{Venturelli2015} and analogically to our previous work  \cite{slysz2022early}, we use the time-indexed JSSP representation. We define binary variables, which encode the starting times of each operation:

\begin{equation}
x_{j,k,t} =
\begin{cases}
             1  & \text{if operation $O_{j,k}$ starts at time $t$} \\
             0  & \text{otherwise}
       \end{cases}
\end{equation}

Since there is no other way to encode these constraints, it is necessary to make them a part of the cost matrix by which we will reward the solver for finding a solution that meets these requirements and punish it for violating the constraints.
There are three hard constraints for JSSP, as well as the optimization objective:
\begin{itemize}
    \item \textbf{Single-start constraint $H_1(x)$}: Each job should start once and only once.
    This constraint ensures, that each operation from each job has exactly one starting time.
    \item \textbf{Machine sharing constraint $H_2(x)$}: At a given time no two jobs should be running on the same machine.
    \item \textbf{Precedence constraint $H_3(x)$}: The precedence of operations within jobs should be maintained.
    This ensures that no operation with a lower index within the same job starts, before the previous one has finished.
    \item \textbf{Minimal makespan constraint $H_4(x)$}:  Promotes low-makespan schedules by putting a penalty on any non-optimal schedule (schedule with finish time further away from the maximum time $T_{\max}$).
\end{itemize}

All those constraints are encoded into the $Q$ matrix coefficients in a form of a weighted sum as we iterate over pairs of variables. The $Q$ matrix can be denoted as:
\begin{equation}
    Q = \sum_{i=1}^4 w_i \cdot H_i \left( x \right),
\end{equation}
\noindent where $w_i$ are weights for respective constraints $H_i(x)$ for $i = 1, 2, 3, 4$.

Also a regularization factor has been added to the objective function. Using the $L2$ regularization in which the number of binary variables equal to $1$ should the same as the total number of operations in all jobs $|O|$. The regularization factor is added with an additional weight $\gamma$. The final optimization cost function takes the form of:

\begin{equation}
\min_{\bm{x}} \bm{x}^T Q \bm{x} + \gamma \left(\sum_i^N x_i - |O|\right)^2.
\end{equation}

\section{Computational experiments}
Within the experimental framework, we wanted to test ORCA PT-1 quantum computers, in selected optimization problems. The quantum machines on which the calculations were run, had $8$ qumodes and allowed running the calculations in two different modes: single-loop mode with $7$ programmable parameters and double-loop mode, with the number of programmable parameters equal to $14$. 

\subsection{Max-Cut}
Max-Cut as a relatively simple combinatorial optimization problem was an excellent benchmark for the quantum device. In order to compare between 5 different experiment configurations we collected statistics on the behaviour of the solution quality, as well as execution time for running the Max-Cut problem with the growing instance size. We run our tests on a quantum device with single-loop and double-loop configurations, as well as simulate those results using a classical simulator and compare it to the optimal solution found by the full enumeration approach.

The size of the problem instance corresponds to the number of binary variables of which the problem consists. In the case of a graph for the Max-Cut problem, the number of graph vertices $|V|$ corresponds to the number of variables. For each number of vertices, connected graphs with different topologies were randomly created. The density of the drawn graph, and therefore the probability of drawing an edge between each pair of vertices, was set to $p=0.8$. This guarantees that the resulting graphs are not too sparse and a given instance will not be trivial. In order to achieve a reasonable payoff between computing time and solution quality we ran $20$ iterations of the BBS algorithm, in each of which we collected $20$ samples. Each experiment was repeated $10$ times.

\begin{figure}[H]
    \centering
    \includegraphics[width=\linewidth]{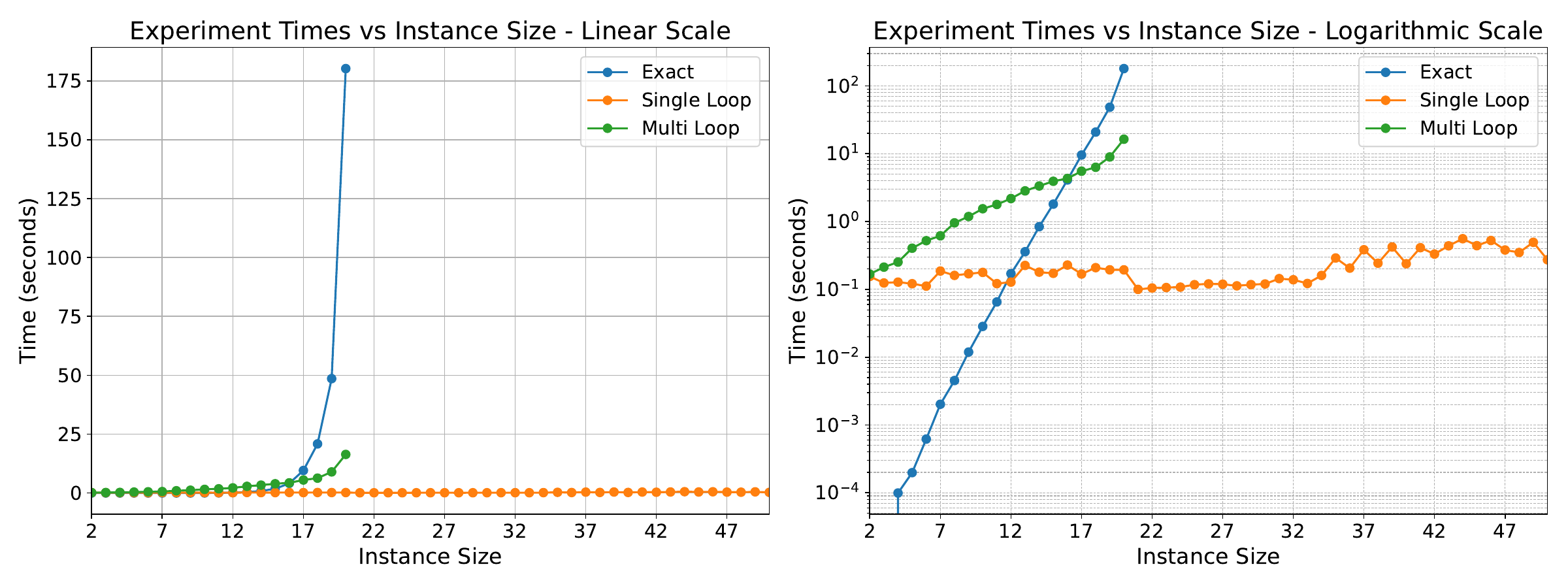}
    \caption{Comparison of computation times between exact search and BBS algorithm simulated classically.}
    \label{fig:simplots}
\end{figure}

The first part of the experiment compares the execution times between an exact solution and the BBS algorithm simulated on a classical device. 
Subplots from Fig. \ref{fig:simplots} show, that the time to find the exact solution explodes quickly, due to the exponential nature of the problem.
Despite the fact that the simulation of the BS process performed by the PT-1 device is also exponential, the heuristic algorithm manages to find the solution in less time, crossing the graph for the exact solution for values as low as a dozen variables ($12$ for single-loop mode and $17$ for double-loop mode).
This, combined with the growth rate of the two graphs, allows us to suggest that the BBS heuristics are much faster than finding the exact solution by brute force algorithm.

The second part of the experiments covers all 5 approaches, including running the BBS algorithm using an actual quantum device. The upper bound for the instance size is determined by the current possibilities of the available BS devices. Another limitation was the number of input photons, due to this fact it was possible to run $2$ and $3$ photon jobs. Fortunately, the possibility of using tiling made it also possible to process larger instances. The input states for each instance size are listed in Table \ref{tab:inputs}, alongside the number of tiles necessary to fit in the instances of a given size. 

\begin{table}[]
\centering
\begin{tabular}{|c|c|c|}
\hline
\textbf{Instance size} & \textbf{Input state}      & \multicolumn{1}{l|}{\textbf{Tilling}} \\ \hline
2                      & {[}1, 0{]}                & 1                                    \\ \hline
3                      & {[}1, 0, 1{]}             & 1                                    \\ \hline
4                      & {[}1, 0, 1, 0{]}          & 1                                    \\ \hline
6                      & {[}1, 0, 1{]}             & 2                                    \\ \hline
8                      & {[}1, 0, 1, 0{]}          & 2                                    \\ \hline
12                     & {[}1, 0, 1{]}             & 4                                    \\ \hline
15                     & {[}1, 0, 1{]}             & 5                                    \\ \hline
20                     & {[}1, 0, 1, 0{]}          & 4                                    \\ \hline
25                     & {[}1, 0, 1, 0, 1{]}       & 5                                    \\ \hline
\end{tabular}
\caption{Different input states, along with the number of tiles, corresponding to the Max-Cut instance size.}
\label{tab:inputs}
\end{table}

\begin{figure}
    \centering
    \includegraphics[width=\linewidth]{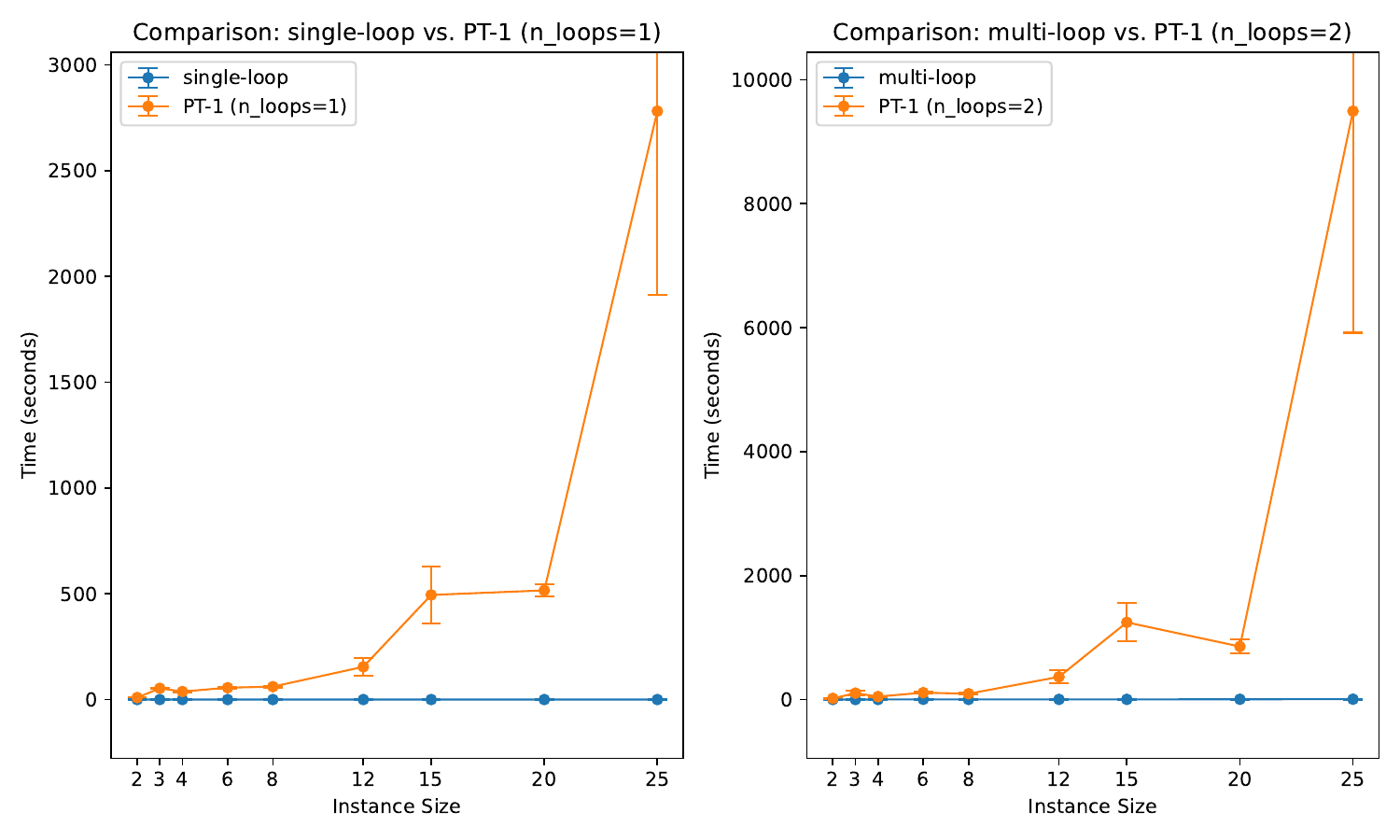}
    \caption{Results for the BBS algorithm on quantum device compared to classical simulation.}
    \label{fig:quantum_vs_classical}
\end{figure}

The plot on Fig. \ref{fig:quantum_vs_classical} shows the comparison between the execution times of full BBS algorithm run on the quantum device, compared to classical simulations for both single and double loop configurations. 

Since we are processing small instances, the execution times of the algorithm, using a quantum device, are much larger than classical ones. The reasons for longer calculation times can vary, but we have identified several. Firstly, the shape of the plot is correlated to the quadratic increase, connected to the tiling parameter. Secondly, the ORCA PT-1 quantum device operates at single Hertz frequencies. This means that only a couple of measurement per second are taken, which for an entire BBS algorithm involving several hundred of such measurements will affect the processing time drastically. Thirdly, the time for communication with the device, as well as calibration, readout and other physical processes is substantial and also has to be taken into consideration.

\begin{figure}[H]
    \centering
    \includegraphics[width=\linewidth]{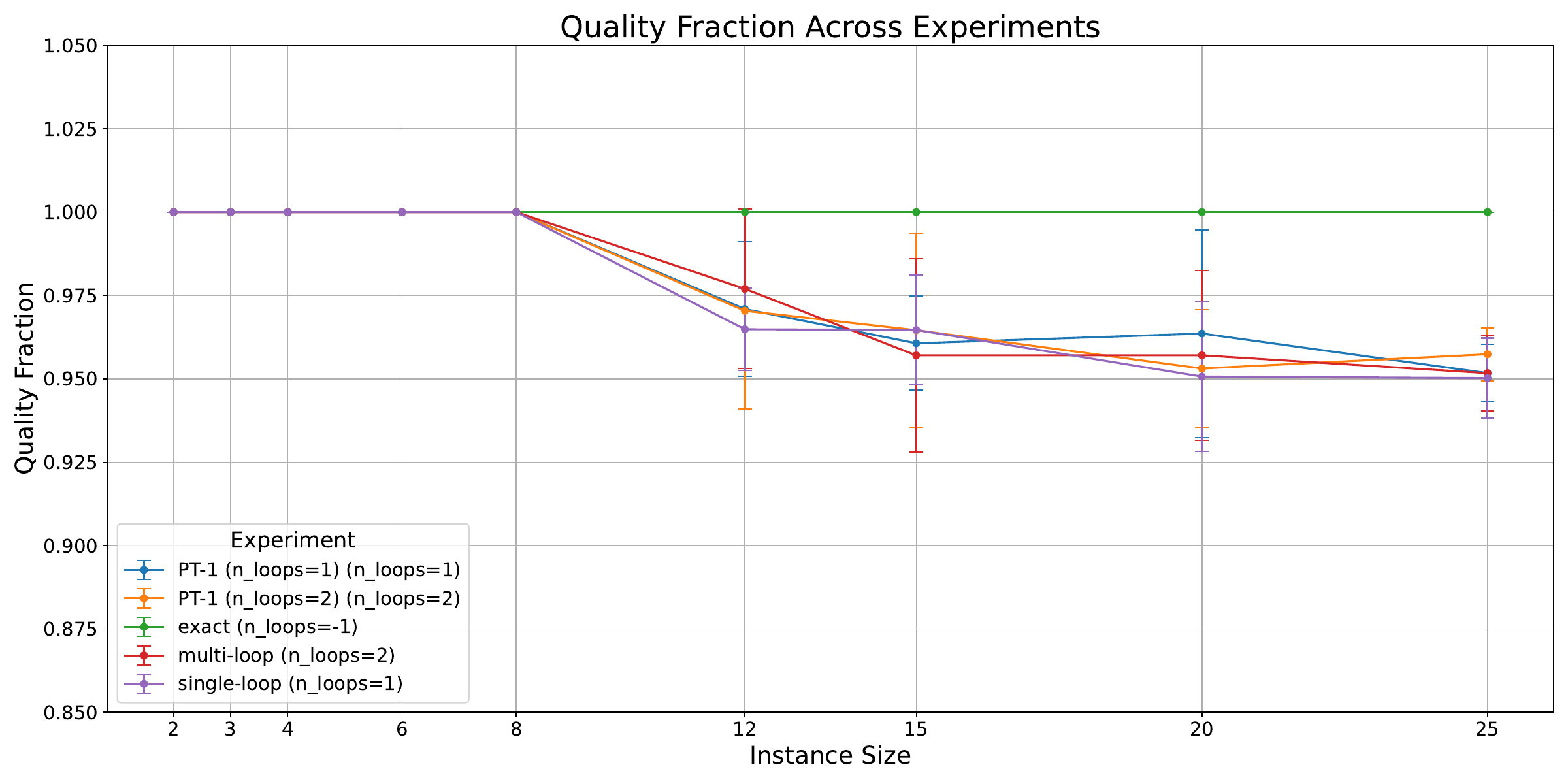}
    \caption{Avarage quality measure for different instance sizes.}
    \label{fig:quality}
\end{figure}

The quality of the result was also measured for each of the 5 experiment configurations. The quality for each Max-Cut instance was defined as the fraction of $\frac{\text{\# cut edges}}{\text{\# cut edges in the exact solution}}$ and could range from 0 to 1. From the plot on Fig. \ref{fig:quality} we can see, that the average quality did not drop below $0.95$, for any instance size. For smaller instances it managed to stay at $1$ and the average quality started to decline slightly for around a dozen variables. A positive observation was that the quality of solutions from the quantum computer did not differ from those from acquired from the simulator, even surpassing it on average in some instances.

\subsection{JSSP}
JSSP is a much more complicated problem to formulate and solve, so our goal was only to run its toy-instance on a photonic quantum device, in order to see if it is possible for the quantum device to process such a complex combinatorial optimization problem. Based on our previous work \cite{slysz2022early}, where we experimented with this problem using a simulator, we compared those results to results of an actual physical quantum computer.

For the experiments we used a toy problem consisting of 3 jobs $\mathcal{J}$ and 2 machines $\mathcal{M}$.

$$\mathcal{J} = \set{ "cupcakes", "smoothie", "lasagna"}$$

$$\mathcal{M} = \set{"mixer", "oven"}$$

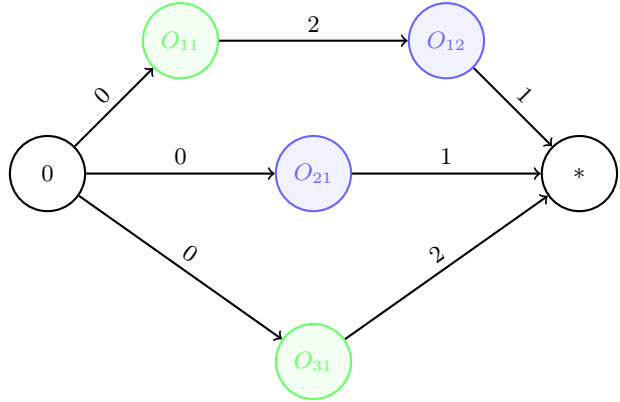
\begin{figure}[H]
    \centering
    \begin{tikzpicture}[node distance={25mm}, thick, main/.style = {draw, circle, minimum size=10mm}, mixer/.style = {draw, circle, minimum size=10mm, color=green!60, fill=green!5,}, oven/.style = {draw, circle, minimum size=10mm, color=blue!60, fill=blue!5}]
    \node[main] (1) {$0$}; 
    \node[mixer] (2) [above right of=1] {$O_{11}$}; 
    \node[oven] (4) [below right of=2] {$O_{21}$}; 
    \node[oven] (5) [above right of=4] {$O_{12}$}; 
    \node[mixer] (6) [below of=4] {$O_{31}$};
    \node[main] (7) [below right of=5] {$ * $}; 
    \draw[->] (1) -- node[midway, above, sloped, pos=0.5] {0} (2); 
    \draw[->] (1) -- node[midway, above, sloped, pos=0.5] {0} (4);
    \draw[->] (1) -- node[midway, above, sloped, pos=0.5] {0} (6);
    \draw[->] (2) -- node[midway, above, sloped, pos=0.5] {2} (5);
    \draw[->] (5) -- node[midway, above, sloped, pos=0.5] {1} (7);
    \draw[->] (4) -- node[midway, above, sloped, pos=0.5] {1} (7);
    \draw[->] (6) -- node[midway, above, sloped, pos=0.5] {2} (7);
    \end{tikzpicture}
    \caption{Example JSSP instance with 3 jobs, 4 operations and 2 machines. $O_{j,k}$ nodes denotes $k$-th operation of job $j$ and colors green and blue correspond to machines \textit{mixer} and \textit{oven} respectively. The numbers on the edges of the graph indicate the processing times of operations labelling preceding vertices.}
    \label{fig:p_graph}
\end{figure}

The complete problem notation also shows execution times of each operation on a given machine, along with the order of operations within jobs is given in a dictionary-like format:\\
\begin{equation}
\begin{split}
\{
"cupcakes": [("mixer", 2), ("oven", 1)],\\
"smoothie": [("mixer", 1)],\\
"lasagna": [("oven", 2)]\},
\end{split}
\nonumber
\end{equation}

\noindent or in a form of a dependency graph as shown in Fig. \ref{fig:p_graph}.

The number of binary variables is calculated as the total number of all operations in all jobs ($|O|$) multiplied by the maximum time constant $T_{\max}$, which has been chosen arbitrarily as the problem size. The only limitation is that $T_{max}$ should not be smaller than the optimal time of a given instance, because if it was, finding a feasible solution would be impossible. In most cases the optimal time is not known, however, one can estimate it based on various factors while preprocessing the instance. For this toy-instance it was easy to find that the optimal makespan was equal to $3$, and hence $T_{max} \ge 3$.

In order to reduce the number of variables, we can perform basic pruning by eliminating variables that, if selected, would generate infeasible solutions. The exclusion of illegal start times is performed, by removing variables that would cause the job to finish after the maximum time or start the operation before the earliest possible time (due to precedence constraints).


 For $T_{max} = 3$ the original number of binary values would be $12$, however, it can be reduced to $7$ after this preliminary preprocessing step as shown in Table \ref{tab:pruning3}.

\begin{table}[H]
\begin{tabular}{|c|c|c|c|}
\hline
\rowcolor[HTML]{34FF34} 
\textbf{Cupcakes - mixer}        & \cellcolor[HTML]{34CDF9}\textbf{Cupcakes - oven} & \textbf{Smoothie - mixer} & \cellcolor[HTML]{34CDF9}\textbf{Lasagna - oven} \\ \hline
$x_{1,1,0}$                         & \cellcolor[HTML]{FD6864}$x_{1,2,0}$                 & $x_{2,1,0}$                  & $x_{3,1,0}$                                        \\ \hline
\cellcolor[HTML]{FD6864}$x_{1,1,1}$                         & \cellcolor[HTML]{FD6864}$x_{1,2,1}$                 & $x_{2,1,1}$                  & $x_{3,1,1}$                                        \\ \hline
\cellcolor[HTML]{FD6864}$x_{1,1,2}$ & $x_{1,2,2}$                                         & $x_{2,1,2}$                  & \cellcolor[HTML]{FD6864}$x_{3,1,2}$                                        \\ \hline
\end{tabular}
\caption{Pruning the variables for simple instance with $T_{max} = 3$. Out of initial $12$ variables ($|O| \times T_{\max}$), $5$ variables marked in red can be pruned.}
\label{tab:pruning3}
\end{table}

\begin{figure}
\includegraphics[width=\textwidth]{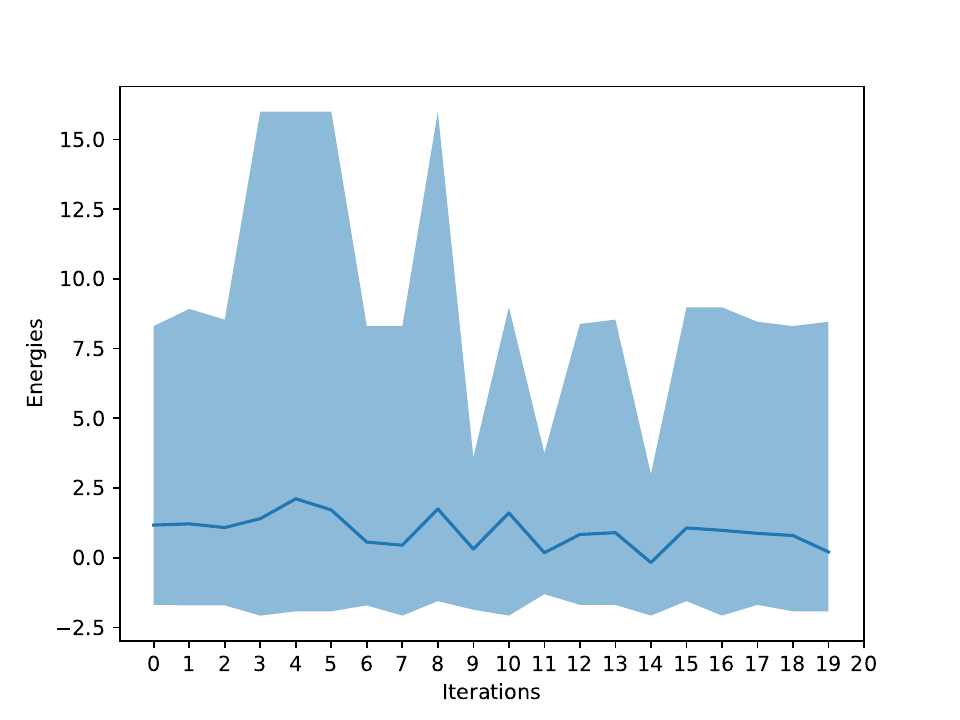}
\caption{Learning curve plot of the BBS hybrid algorithm for the 7 variable instance of the JSSP problem.} \label{fig:learning_curve}
\end{figure}
\begin{figure}
    \centering
    \includegraphics[width=1.1\linewidth]{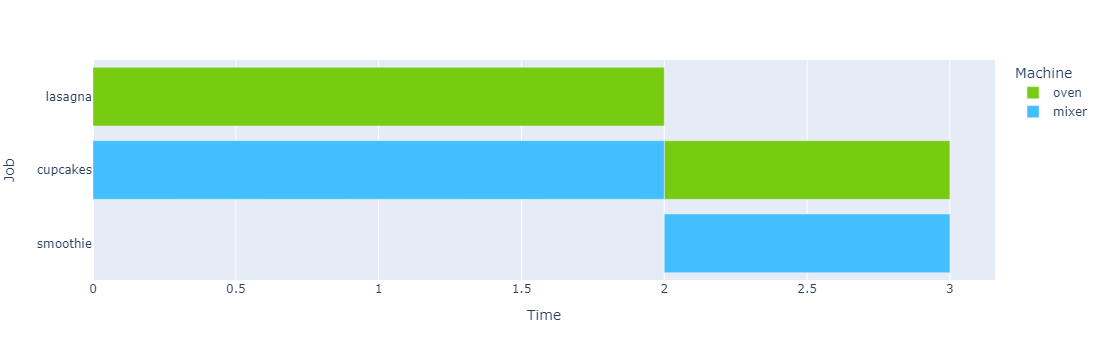}
    \caption{Gantt chart of the optimal solution for the analyzed instance found by the BBS algorithm run on a photonic quantum computer.}
    \label{fig:gantt}
\end{figure}

This 7 variable problem can be solved with a PT-1 device, by utilizing the input state {[}1, 0, 1, 0{]} tiled $2$ times with a padding of $1$. The quantum computer was able to find the optimal solution with constraint weights set to: $w_1 = 1$, $w_2 = 2$, $w_3 = 5$, $w_4 = 1$ and $\gamma = 1$.
The learning process is visualized in Fig. \ref{fig:learning_curve}. The solid line on the graph visualizes the average value of the cost function for the samples from each iteration, and the shaded area indicates the range of cost function values for the entire batch of samples (minimum to maximum values). We can see that the continuous line gradually decreases, which translates into finding better solutions of the samples, which are then returned as optimal.
The Gantt chart of the optimal solution, which for $T_{max} = 3$ is also the only feasible solution, is visualized in Fig. \ref{fig:gantt}.

\section{Conclusions}
In this paper we have described a novel approach to solve combinatorial optimization problems with a photonic quantum computer based on the Boson Sampling paradigm. We covered the advantages and disadvantages of various quantum computing architectures and corresponding optimization algorithms following with benchmarking the photonic quantum system on the example of two optimization problems: Max-Cut and JSSP  to measure computation time and solution quality. For our research we used an actual 8-qumode ORCA PT-1 photonic quantum system, installed on-site at Poznań Supercomputing and Networking Center.

Using the example of the Max-Cut problem for the instances that were run, it was possible to show that the BBS algorithm, that used simulations of a quantum device, is much faster than finding an exact solution. It was also possible to run the BBS algorithm on an actual quantum device, proving that they can work for instances as big as $25$ variables. Although execution times were much higher, we were able to identify the most important reasons for this behavior. Importantly, each of them can be addressed in subsequent hardware and software upgrades of the device, and the execution time of the entire algorithm only increases polynomially with a growing instance size.

The quality of solutions returned by the BBS algorithm was very satisfactory, remaining above 95\% of optimality for all analyzed instances. In addition, it can be noted that the results do not differ significantly between the simulator and the real quantum computer, which leads to believe that when the capabilities of quantum computers are improved, it will be possible to solve larger instances of such optimization problems very efficiently.

The Job-Shop Scheduling Problem was much more difficult to solve for the algorithm, as its structure is more complex. Not only does it have a cost function but also a number of complex constraints. Nevertheless, it was possible to repeat the experiments from our previous work and obtain the optimal solution for a toy-instance, this time on a real photonic quantum processor. This shows, that the BBS algorithm can handle more complicated optimization problems and can be successfully utilised on an actual quantum device.

The nature of this work is a demonstration of the capabilities of real quantum machines installed on-site in a non-specialized facility, which for many reasons is already a huge advantage over running quantum computations via cloud. Of course, the results of the optimization experiments are only preliminary, but they show the potential and capabilities of these machines in actual applications such as combinatorial optimization. With further development of photonic quantum computers, it may be possible to surpass the encountered challanges and limitations. With this progress, achieving quantum advantage with photonic quantum computers can be seen a realistic goal.

Future work associated with optimization algorithm on photonic quantum systems may include solving some of the previously mentioned challenges, as well as scaling and speeding up the computation to a level where it is possible to surpass classical simulation time and quality. Such progress would require work with both software and hardware components and cannot be done in isolation. However, due to the procurement of smaller and easier-to-maintain quantum systems, this is a likely route, as such an investment model allows for closer collaboration within the community and can significantly speed up the further research process.\\

\textbf{Acknowledgments}\\

This research has been funded by the Program of the Polish Ministry of Science and Higher Education "Applied Doctorate" realized in years 2022-2026 (agreement no. DWD/6/0142/2022) and by the Poznan University of Technology (project no. 0311/SBAD/0746).

\clearpage
\bibliographystyle{splncs04}
\bibliography{bibliography}

\end{document}